\documentclass[copyright,creativecommons]{eptcs}
\providecommand{\event}{TERMGRAPH 2013}

\providecommand{\firstpage}{1}

\usepackage{breakurl}

\usepackage{subfig}
\usepackage{amsmath,amssymb,stmaryrd}
\usepackage{xspace}
\usepackage{hyperref, graphicx}
\usepackage{pstricks}

\usepackage{macros,nets}

\title{Bigraphical Nets}
\author{Maribel Fern\'andez
\and Ian Mackie 
\and Matthew Walker
\institute{\'Ecole Normale Sup\'erieure,  Paris, France\\
\'Ecole Polytechnique,  Palaiseau, France\\
King's College London, Dept.\ of Informatics,  London WC2R 2LS, UK}
}

\def\authorrunning{M. Fern\'andez, I. Mackie \& M. Walker}
\def\titlerunning{Binets}

\begin{document}
\maketitle

\begin{abstract} 
Interaction nets are a graphical model of computation, which has been
used to define efficient evaluators for functional calculi, and
specifically $\lambda$-calculi with patterns.  However, the flat
structure of interaction nets forces pattern matching and functional
behaviour to be encoded at the same level, losing some potential
parallelism.  In this paper, we introduce bigraphical nets, or
binets for short, as a generalisation
of interaction nets using ideas from bigraphs and port graphs, and we
present a formal notation and operational semantics for binets.  We
illustrate their expressive power by examples of applications.

\end{abstract}

\section{Introduction}

Interaction nets~\cite{LafontY:intn} are graphical rewrite systems
used for the specification of logical proof systems
(e.g.,~\cite{AlvesS:gracp,LafontY:fropni}), for the implementation of
efficient evaluators for the $\lambda$-calculus
(e.g.,~\cite{GonthierG:geoolr,AspertiA:bolhom,MackieIC:efflei}), and
for visual programming
(e.g.,~\cite{MackieIC:diagrams,MackieIC:tamc,MackieIC:vcc}).

The visual nature of interaction nets makes them well suited as a
specification tool, and, since \emph{all} the computation steps are
explicit and expressed in the same formalism (there is no external
machinery), interaction nets are also well suited for the study of the
dynamics of programming languages and rewriting
systems~\cite{MackieIC:intntr,FernandezM:inMCpa,PintoJS:seqcam}. 
However, interaction nets have some drawbacks. 
When the nets are large or growing during reduction, being able to
\emph{structure} the graph is crucial to understand the system
modelled, but interaction nets lack mechanisms to structure the system.
Moreover, to formally prove properties of the system modelled or
implement reduction, a \emph{formal, algebraic notation}, with a
precise operational semantics, should also be available. In this
paper, we address these two points:
\begin{itemize}
\item
 First, inspired by Milner's bigraphs~\cite{Milner}, we define a
 generalisation of interaction nets, which we call bigraphical nets,
 or simply \emph{binets}, where not only the connectivity but also
 the hierarchical structure of the system is taken into
 account. Binets borrow from bigraphs a notion of locality that is
 missing in interaction nets.
\item
Then,  we present a
formal algebraic notation for binets, with an operational semantics
which can serve as a basis for their implementation.
\end{itemize}

\paragraph{Related Work.}
Binets can be seen as hierarchical graph rewriting systems that permit
links between nested nets and external subgraphs (like bigraphs, and
unlike hierarchical graphs~\cite{Drewes-Hoffmann-Plump:00}). Rewriting
can take place across boundaries.  Both of these features will be of
use in our encoding of the $\rho$-calculus.

Binets inherit from interaction nets the notion of principal
port. But, in contrast with interaction nets, binets do not force all
interactions to be binary, and in contrast with bigraphs, they place
restrictions on reactions to simplify the implementation of rule application.

Interaction nets have been used as an implementation language for
functional calculi, and as a tool to understand their
dynamics~\cite{GonthierG:geoolr,AspertiA:bolhom,MackieIC:phd,MackieIC:efflei,FernandezM:terrgi,FernandezM:inMCpa,FleutotF:encoci}.
Interaction net encodings of the
$\rho$-calculus~\cite{rhoCalIGLP-I+II-2001}, an extension of the
$\lambda$-calculus where we can abstract on patterns, not just on
variables, shed light on the implicit parallelism present in the
$\rho$-calculus, and at the same time, motivate a generalisation of
interaction nets, as shown in~\cite{EXPRESS}.  In this paper, we
develop and formalise this idea. Our main contribution is a formal
syntax and operational semantics for binets, via a textual calculus.

The class of portgraphs defined by Andrei and Kirchner~\cite{AndreiK08}
can also be seen as a generalisation of interaction nets, but although
binets are graphs with ports, due to their hierarchical nature they
cannot be defined as portgraphs. It would be interesting to consider a
generalisation of portgraphs with a notion of locality; the inclusion
of this feature in PORGY~\cite{AndreiO:PORGY} could serve as a
starting point for the development of a specification environment
based on binets.


\section{Background}
\label{sec:bg}

\paragraph{Interaction Nets.}
A system of interaction nets is specified by a set $\Sigma$ of symbols
with fixed arities, and a set $\IR$ of interaction rules.  An
occurrence of a symbol $\alpha\in\Sigma$ is called an \emph{agent}. If
the arity of $\alpha$ is $n$, then the agent has $n+1$ \emph{ports}: a
\emph{principal port} depicted by an arrow, and $n$ \emph{auxiliary
ports}. Such an agent will be drawn in the following way:
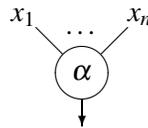
\begin{net}{40}{40}
\putalpha{20}{20}
\putDvector{20}{10}{10}
\putline{12.6}{27.4}{-1}{1}{10}
\putline{27.4}{27.4}{1}{1}{10}
\puttext{20}{35}{$\cdots$}
\put(2.6,38){\makebox(0,0)[br]{$x_1$}}
\put(37.4,38){\makebox(0,0)[bl]{$x_n$}}
\end{net}
A net $N$ is a graph (not necessarily connected) with agents at the
vertices and each edge connecting at most 2 ports. The ports that are
not connected to another agent are \emph{free}. There are two special
instances of a net: an empty net, and a net consisting only of edges
(no agents).  The \emph{interface} of a net is the set of free ports
of agents and free extremes of wires. We refer to~\cite{LafontY:intn}
for more details.

An interaction rule $((\alpha,\beta) \Lra N) \in \IR$ replaces a pair
of agents $(\alpha,\beta)\in \Sigma\times\Sigma$ connected together on
their principal ports (an \emph{active pair} or \emph{redex} and
written $\alpha \bowtie \beta$) by a net $N$ with the same
interface. Rules must satisfy two conditions: all free ports are
preserved during reduction (there are no global operations: only the part of
the net involved in the rewrite is modified), and there is at most one
rule for each pair of agents (such a rule will thus be
sometimes denoted by $\alpha \bowtie \beta$). The following
diagram shows the format of interaction rules ($N$ can be any net
built from $\Sigma$).

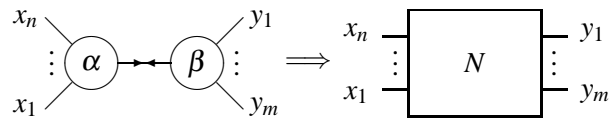
\begin{net}{200}{60}
 \putalpha{20}{20} 
 \putbeta{60}{20} \putRvector{30}{20}{10}
 \putLvector{50}{20}{10} \putline{12.6}{27.4}{-1}{1}{10}
 \putline{12.6}{12.6}{-1}{-1}{10} \putline{67.4}{27.4}{1}{1}{10}
 \putline{67.4}{12.6}{1}{-1}{10} \puttext{5}{23}{$\vdots$}
 \puttext{75}{23}{$\vdots$} \put(0,0){\makebox(0,0)[br]{$x_1$}}
 \put(0,40){\makebox(0,0)[tr]{$x_n$}}
 \put(80,0){\makebox(0,0)[bl]{$y_m$}}
 \put(80,40){\makebox(0,0)[tl]{$y_1$}} \puttext{102}{20}{$\Lra$}
 \putbox{140}{0}{50}{40}{$N$} \putHline{130}{10}{10}
 \putHline{130}{30}{10} \puttext{135}{23}{$\vdots$}
 \putHline{190}{10}{10} \putHline{190}{30}{10}
 \puttext{195}{23}{$\vdots$} \put(125,5){\makebox(0,0)[br]{$x_1$}}
 \put(125,35){\makebox(0,0)[tr]{$x_n$}}
 \put(205,5){\makebox(0,0)[bl]{$y_m$}}
 \put(205,35){\makebox(0,0)[tl]{$y_1$}} 
\end{net}

We use the notation $\Lra$ for the one-step reduction relation, or
$\Lrawith{\alpha \bowtie \beta}$ if we want to be explicit about the
rule used, and $\Lra^*$ for its transitive and reflexive closure. If a
net does not contain any active pairs then we say that it is in normal
form. The key property of interaction nets is that reduction is
strongly confluent.  We refer the reader to~\cite{LafontY:intn} for
more details and examples.

\paragraph{Bigraphs.}
In~\cite{Milner,jensen03bigraphs} a notion of graph transformation
system is defined, using nested (or hierarchical) graphs called
\emph{bigraphs}.  Bigraphs represent two kinds of structure: locality
(nodes may occur inside other nodes) and connectivity (nodes have
ports that may be connected by links). We recall the basic terminology
of bigraphs and refer the reader to~\cite{Milner} for details and
examples.

A bigraph is a pair of a \emph{place graph} and a \emph{link graph}
over the same set of nodes. 
It has interfaces, which define the way in which it can be composed
with other bigraphs. The place graph, or placing, is a set of trees
with interfaces, and the link graph, or linking, is a hypergraph with
interfaces. A placing has inner and outer interfaces. The inner
interface corresponds to the \emph{sites} where other graphs can be
placed, and the outer interface corresponds to the \emph{roots} of the
trees. The linking also has inner and outer interfaces, which are
names of ports, that is, the points where edges can be attached.

Nodes are labelled by \emph{controls} with fixed arities; the arity of
a control corresponds to the number of ports of the node.  A control
is \emph{atomic} if it cannot contain a nested graph, otherwise it is
non-atomic.

The reduction relation is defined by a set of reaction rules, which
are pairs of bigraphs (called redex and reactum). The redex has a
\emph{width}, corresponding to the number of sites it occupies in the
outer bigraph \cite{jensen03bigraphs}.  A non-atomic control $K$ can
be specified as active, in which case reactions can occur inside, or
passive, in which case reactions in the internal bigraph can only
occur after the control $K$ has been destroyed.

Interaction nets can be seen as a particular kind of bigraphs without
nesting: all controls (called agents in interaction nets) are atomic,
and have a distinguished port (the principal port). Interaction rules
can be seen as reactions in which both redex and reactum have width 1,
and redexes are restricted to just two controls connected by one link
through the distinguished ports.

\section{Binets}
\label{sec:binets}

\subsection{Informal presentation}

Bigraphs~\cite{jensen03bigraphs} introduce a notion of locality (using
nesting to indicate that a graph is local to a certain node) which is
missing in interaction nets. In this section, we define binets as a
generalisation of interaction nets to incorporate this feature.
We start with an informal definition of binets, contrasting them with interaction nets,
before presenting a formal syntax and semantics for them.

A binet is a labelled graph consisting of a set of nodes (also called
\emph{agents}) and a set of edges, which are attached to nodes at
connection points called \emph{ports}.  Each edge connects at most two
ports.  The label of a node (i.e., the agent's name) determines its
arity, that is, the number of ports it has. Each agent has a
distinguished port, called the \emph{principal} port, and a (possibly
empty) set of \emph{auxiliary} ports.  An agent can be located inside
another agent, and edges can connect ports of agents situated at
different nesting levels (i.e., edges can cross node boundaries).

\emph{Interaction rules}, also called reaction rules, define interactions
between two agents connected by their principal ports, 
 or interactions of an agent with its local subnets, preserving
the interfaces.

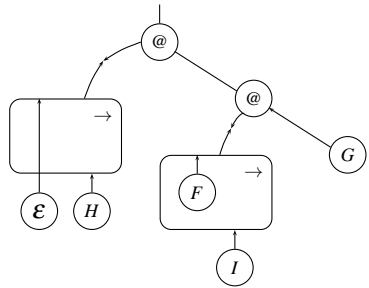
\begin{figure}
\begin{center}
\scalebox{0.5}{
\begin{pspicture}[showgrid=false](-12,-1)(28,6.5)

\psframe[framearc=0.3](0,4)(3,2)
\rput(2.5,3.5){\Large{$\rightarrow$}}
\pscurve{->}(2,4)(2.2,4.5)(2.5,5)
\pscurve{<-}(2.5,5)(3,5.3)(3.5,5.5)

\pscircle(4,5.5){.5}
\psline(4,6)(4,6.5)
\rput(4,5.5){\Large{$@$}}

\pscircle(6.5,4){.5}
\psline(6.05,4.2)(4.4,5.25)
\rput(6.5,4){\Large{$@$}}

\pscircle(9,2.5){.5}
\psline{->}(8.55,2.7)(6.9,3.75)
\rput(9,2.5){\Large{$G$}}

\psframe[framearc=0.3](4,0.5)(7,2.5)
\rput(6.5,2){\Large{$\rightarrow$}}
\pscurve{->}(5.6,2.5)(5.7,2.8)(5.9,3.2)
\pscurve{<-}(5.9,3.2)(6.2,3.6)(6.25,3.6)

\pscircle(5,1.5){.5}
\psline{->}(5,2)(5,2.5)
\rput(5,1.5){\Large{$F$}}

\pscircle(6,-.5){.5}
\psline{->}(6,0)(6,0.5)
\rput(6,-.5){\Large{$I$}}

\pscircle(0.8,1){.5}
\psline{->}(0.8,1.5)(0.8,4)
\rput(0.8,1){\huge{$\epsilon$}}

\pscircle(2.2,1){.5}
\psline{->}(2.2,1.5)(2.2,2)
\rput(2.2,1){\Large{$H$}}
\end{pspicture}
}
\end{center}
\caption{Binet for the $\rho$-term $(x \rightarrow H) ((F \rightarrow I)G)$.}
\label{fig:binet-ex}
\end{figure}

Figure~\ref{fig:binet-ex} shows a binet representing a $\rho$-term.
Ovals and circles represent agents, their names are written inside;
principal ports are marked with an arrow, the free port at the top is
marked by a dangling edge. The $\epsilon$-agent is drawn outside the
$\rightarrow$-agent to exploit a non-strict semantics as early in the
reduction process as possible.

In contrast with interaction nets, the left-hand side of a reaction
rule can specify the location in which the reacting agents are, or
the locations contained in these controls, and reactions can take
place across boundaries.  However, reduction is still local in the
sense that it only affects the nodes that match the left hand side (no
global conditions or updates are specified).  The latter point is
relevant for implementation.  

Agents in binets correspond to the
notion of control in bigraphs, and binet reaction rules are a particular
class of bigraph reaction rules. Each binet 
has an associated place graph and link graph, similarly to bigraphs.
All the examples of bigraphs for the $\pi$-calculus and
ambient calculus given in~\cite{jensen03bigraphs} (part I) can be
recast as binets by adding principal ports and copy/erase agents (controls)
to preserve the interface of the reactions.

Comparing with the properties of interaction nets, we remark that
confluence does not hold in general for binets, because of the
possibility of interactions across boundaries. 
To study the formal properties of binet reduction, below we give a 
calculus for binets.

\subsection{A calculus for binets}
\label{sec:formal}

As Milner~\cite{Milner} stated: ``Diagrams are valuable for rapid
appreciation of a system's structure. On the other hand, algebra is
essential to express [...] the ways in which a system may be resolved
into components.'' In this section we give a formal, algebraic
presentation for binets.  First we give the syntax of the language,
and then we present an operational semantics for programs written in
this language.

\paragraph{Syntax.}
\newcommand\bnet[5]{\ensuremath{#1^{#2} \langle #3 \,|\, #4 \,|\, #5 \rangle}}
\newcommand\bnets[2]{\ensuremath{#1^{#2} \langle \rangle}}
\newcommand\bnetm[3]{\ensuremath{#1^{#2} \langle #3 \rangle}}

A textual syntax for binets has to capture, dually, the connections
between agents (including where those connections are principal
ports), differentiating between internal or external with respect to
the originating agent and also the locality of agents within a system,
i.e., their physical position within other agents. It is the intention
of this syntax to unambiguously state these three properties without
over-complication. 

We define below agents and binets over a \emph{signature}
$\mathcal{A},L$, where $\mathcal{A}$ is a set of \emph{agent names},
each with an associated arity $(n,m)$ corresponding to the number of
ports in its internal and external interface, respectively, and $L$ is
a set of port labels. We assume $L \cap \mathcal{A} = \emptyset$.

\begin{definition}[Agent]
An \emph{agent} over the signature $\mathcal{A}$, $L$ is written
$\bnet{A}{l}{E}{I}{N}$, where $A\in\mathcal{A}$ is the agent name,
which determines its arity $(n,m)$, $l \in L$ is the label given to
the principal port of $A$, and the lists $I, E$ of lengths $n,m$
respectively, whose elements are port labels in $L$, denote the
internal and external agent interfaces; the order denotes the
geographic position of the ports (in similar fashion to interaction
nets reading in a clockwise direction from the principal port for
external ports and, without loss of precision, in an arbitrary
clockwise direction for interior ports). $N$ is a (possibly empty)
list corresponding to the set of agents located within the agent.
\end{definition}
The definition above is inductive (due to the inclusion of the set $N$
within $A$) but not recursive; agents are not permitted to be located within themselves.

\begin{definition}[Binet]
A binet over the signature $\mathcal{A}$, $L$ is defined as a set of
agents and wires over $\mathcal{A}$, $L$. Agents have already been
defined. A \emph{wire} is an edge joining two ports, written $a\!-\!b$
where $a,b$ are the labels of the ports.  Each port label in $L$
occurs at most twice in a binet; the \emph{net interface} of the binet
is defined as the subset of labels occurring only once.

The binet containing no
agents and wires is a special case.  For brevity, agents of the form
$\bnet{A}{i}{X}{\emptyset}{\emptyset}$ will be denoted
$\bnetm{A}{i}{X}$ and the particular case when $X = \emptyset$
will be written $\bnets{A}{i}$.
\end{definition}

 A binet, similar to a bigraph, can be decomposed into a place
graph and a link graph. The link graph is explicit in the definition of
binet (a binet is a set of agents and wires); the place graph can 
be reconstructed from the nesting of agents.

Reduction in binets occurs on \emph{active pairs}, which are pairs of
agents connected via their principal ports, similarly to interaction
nets although significantly rules in binets must be aware of locality
context. More precisely, a rule may affect the agents in the place
graph of the active pair.  We illustrate it with an example:
the interaction rule presented in~\cite{EXPRESS} between the matching agent,
 $M$ and any other agent $\alpha$
 is one such occurrence and would be written:
$$\bnet{M}{a}{b}{\emptyset}{X}
\, , \,\bnetm{\alpha}{a}{Y} \Rightarrow \bnet{\alpha_M}{b}{Y}{\emptyset}{X}$$
The metavariables $X$ and $Y$ denote a subnet and series of labels respectively
that remain unchanged under graph reduction.  Here, the interaction between the
agents $M$ and $\alpha$ causes the subgraph located within the agent $M$ (denoted by $X$) to 
move to the new agent named $\alpha_M$ with principal port $b$.

Contrary to interaction nets, binets permit reductions to occur on certain 
graph configurations despite the nonexistence of an active pair, called 
$inactive$ rewriting in the sequel. See for example the 
configuration in \cite{EXPRESS} of an empty matching agent, $M$, where the net is rewritten to a wire without interaction through active pairs.
This is written as follows (the arrow explicitly shows the type 
of rule being applied):
$$\bnetm{M}{a}{b} \Longrightarrow_{inactive} a\!-\!b$$

The reduction calculus is defined below, but first we give an example:
a  reduction sequence for the binet shown in the
previous subsection, representing the $\rho$-term $(x \rightarrow H)
((F \rightarrow I)G)$. The textual representation of each binet is
shown in the table below.

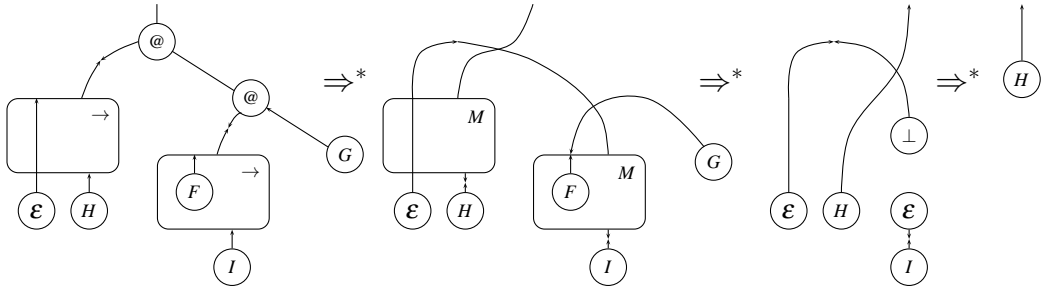
\begin{center}
\scalebox{0.5}{
\begin{pspicture}[showgrid=false](0,-1)(28,6.5)

\psframe[framearc=0.3](0,4)(3,2)
\rput(2.5,3.5){\Large{$\rightarrow$}}
\pscurve{->}(2,4)(2.2,4.5)(2.5,5)
\pscurve{<-}(2.5,5)(3,5.3)(3.5,5.5)

\pscircle(4,5.5){.5}
\psline(4,6)(4,6.5)
\rput(4,5.5){\Large{$@$}}

\pscircle(6.5,4){.5}
\psline(6.05,4.2)(4.4,5.25)
\rput(6.5,4){\Large{$@$}}

\pscircle(9,2.5){.5}
\psline{->}(8.55,2.7)(6.9,3.75)
\rput(9,2.5){\Large{$G$}}

\psframe[framearc=0.3](4,0.5)(7,2.5)
\rput(6.5,2){\Large{$\rightarrow$}}
\pscurve{->}(5.6,2.5)(5.7,2.8)(5.9,3.2)
\pscurve{<-}(5.9,3.2)(6.2,3.6)(6.25,3.6)

\pscircle(5,1.5){.5}
\psline{->}(5,2)(5,2.5)
\rput(5,1.5){\Large{$F$}}

\pscircle(6,-.5){.5}
\psline{->}(6,0)(6,0.5)
\rput(6,-.5){\Large{$I$}}

\pscircle(0.8,1){.5}
\psline{->}(0.8,1.5)(0.8,4)
\rput(0.8,1){\huge{$\epsilon$}}

\pscircle(2.2,1){.5}
\psline{->}(2.2,1.5)(2.2,2)
\rput(2.2,1){\Large{$H$}}

\rput(9,4.5){\Huge{$\Rightarrow^*$}}

\psframe[framearc=0.3](10,4)(13,2)
\rput(12.5,3.5){\Large{$M$}}
\pscurve{}(12,4)(12.3,5)(13.5,5.5)(14,6.5)
\psline{->}(12.2,2)(12.2,1.75)

\pscircle(18.8,2.3){.5}
\pscurve{->}(18.55,2.7)(17,4)(15.5,3.7)(15,2.5)
\rput(18.8,2.3){\Large{$G$}}

\psframe[framearc=0.3](14,0.5)(17,2.5)
\rput(16.5,2){\Large{$M$}}
\pscurve{}(16,2.5)(15.8,3.5)(14,5)(12,5.5)
\psline{->}(16,.5)(16,.25)

\pscircle(15,1.5){.5}
\psline{->}(15,2)(15,2.5)
\rput(15,1.5){\Large{$F$}}

\pscircle(16,-.5){.5}
\psline{->}(16,0)(16,0.25)
\rput(16,-.5){\Large{$I$}}

\pscircle(10.8,1){.5}
\psline{}(10.8,1.5)(10.8,4)
\pscurve{->}(10.8,4)(11,5.2)(12,5.5)
\rput(10.8,1){\huge{$\epsilon$}}

\pscircle(12.2,1){.5}
\psline{->}(12.2,1.5)(12.2,1.75)
\rput(12.2,1){\Large{$H$}}

\rput(19,4.5){\Huge{$\Rightarrow^*$}}


\pscircle(24,-.5){.5}
\psline{->}(24,0)(24,0.25)
\rput(24,-.5){\Large{$I$}}

\pscircle(24,1){.5}
\psline{->}(24,0.5)(24,0.25)
\rput(24,1){\huge{$\epsilon$}}

\pscircle(24,3){.5}
\pscurve{->}(24,3.5)(23.8,4.5)(23,5.3)(22,5.5)
\rput(24,3){\Large{$\bot$}}

\pscircle(20.8,1){.5}
\psline{}(20.8,1.5)(20.8,4)
\pscurve{->}(20.8,4)(21,5.2)(22,5.5)
\rput(20.8,1){\huge{$\epsilon$}}

\pscircle(22.2,1){.5}
\pscurve{->}(22.2,1.5)(22.4,3)(23.8,5)(24,6.5)
\rput(22.2,1){\Large{$H$}}

\rput(25.3,4.5){\Huge{$\Rightarrow^*$}}

\pscircle(27,4.5){.5}
\psline{->}(27,5)(27,6.5)
\rput(27,4.5){\Large{$H$}}

\end{pspicture}
}
\end{center}

\begin{center} 
\begin{tabular}{llll}
First Binet & Second Binet & Third Binet & Fully Reduced
\\
\hline 
\bnet{\rightarrow\!}{x}{a}{b}{\emptyset} & \bnetm{M}{a}{c} & \bnets{H}{c}
& \bnets{H}{c}\\ \bnets{\epsilon}{b} & \bnets{H}{a} &  &\\
\bnets{H}{a} &  & \bnets{\bot}{d} &\\
\bnetm{@}{x}{c,d} & \bnets{\epsilon}{d} & \bnets{\epsilon}{d} &\\
\bnetm{@}{y}{d,e} & \bnet{M}{f}{d}{\emptyset}{\bnets{F}{e}} &
 &\\ \bnet{\rightarrow\!}{y}{f}{g}{\bnets{F}{g}} &
\bnets{I}{f} & \bnets{I}{f} &\\ \bnets{I}{f} & \bnets{G}{e} &
\bnets{\epsilon}{f} &\\ \bnets{G}{e} & & &\\
\end{tabular}
\end{center}

The second binet contains three active pairs but parallel
firing of the rule for agents $M^f$ and $I^f$ with the rule for $F^e$
and $G^e$ would clearly be incorrect; the general rule for $M$ with
$\alpha$ involves a rewriting that affects all of the nested nets
within $M$, hence a reduction strategy is required.  Informally, there
is a choice to delay either active pair until the other has had the
opportunity to reduce (although in this instance either derivation
will eventually lead to the intended destruction of this disjoint
net). A strategy is also required in the same configuration to either
fire the rule for $M^a$ and $H^a$ or, as in the example, to perform an
inactive rewrite on $M^a$ and reduce it to the wire $a-c$ (see the
second textual rule above).

\paragraph{Reduction Rules.}
Rewrite rules follow the simplicity of interaction nets when the
reductum has no rewrite implications for any agent except for the
agent (in the case of inactive rewriting) or agents (active pair
  rewriting) directly involved.  However, due to the more expressive
graph rewriting allowed by binets, rewrite rules require additional
machinery to resolve rewriting of nested agents within the place graph
of the net. Unlike interaction net rules they incorporate
metavariables and an additional strategy language.

A priori knowledge of how a rule may affect the surrounding subnet is essential when the rule is defined.  For example, the $\epsilon$ rule within the $\rho$-calculus scheme is the garbage collection agent responsible for deleting nets.  This agent propagates through subnets, terminating when it forms an active pair with another $\epsilon$-agent, hence when defining the interaction of $\epsilon$ and $M$, the subnets within $M$ should also be reduced, as follows:
$$
\bnets{\epsilon}{a}, \, \bnet{M}{a}{b}{\emptyset}{X} \Longrightarrow
\bnets{\epsilon}{b},\,X,\,
\mbox{foreach }x\mbox{ in }I(X)\mbox{: }\bnets{\epsilon}{x}\mbox{ and }\bnets{\epsilon}{\bar{x}}
$$ 
where $I(X)$ is the collection of labelled ports that constitute
the interface of the nested net $X$ and $\bar{x}$ is a fresh label for
the port outside of $M$ that was connected to the interface at $x$.

The ability of binets to rewrite over agent boundaries means the
efficiency (measured in the number of interactions: typical of
interaction nets although cruder for binets) can be improved by
rephrasing the rule to propagate $\epsilon$ only over the wires that
are free in this subnet (i.e., those wires that extend beyond the
locality of the $M$-agent).  Each of these wires can be identified by a
label appearing only once in the subnet $X$ and so the above rule can
be reinterpreted as follows, where $X$ is removed in one step:
$$
\bnets{\epsilon}{a}, \, \bnet{M}{a}{b}{\emptyset}{X} \Longrightarrow
\bnets{\epsilon}{b},\,
\mbox{foreach }x\mbox{ in }L_X\mbox{ where }x\mbox{ is unique:
}\bnets{\epsilon}{x} 
$$
where $L_X$ is the multiset of labelled ports in $X$.

The strategy language is left informal at this stage with full details
to be provided in later technical reports.

\paragraph{Reduction Calculus.}
The reduction calculus comprises four main parts: firstly populating
a set of the active pairs within a binet and also those (sub)nets that
are configured in a way that permits inactive rewriting.
This collection of active pairs and nets is then prioritised according
to a given reduction strategy 
and, crucially, a collection of agents and nets that can safely be
rewritten in parallel is identified.  Then, both active and
inactive rewriting is performed and, lastly, a tidying stage is
performed to eliminate every explicitly written wire in the net.

\begin{itemize}
\item[Collect] Let $\mathcal{C}_A$ be the set of labels of principal
  ports involved in active pairs (these are easily computed: scan the
  graph and identify each label $l$ that appears twice as a principal
  port) and $\mathcal{C}_I$ be the set of binets that are isomorphic
  to the left hand side of inactive rewriting rules (computed using
  standard subgraph matching algorithms that can be optimised due
  to occurrence of principal ports).
\item[Prioritise] According to the reduction strategy implemented
  (weighted, stochastic, typed and so on) group all \emph{safe} nets,
  $\mathcal{C}_s$.  A collection of safe nets is one where rewriting
  (either active or inactive) can occur in parallel without conflict.
  The safety, or otherwise, of potential net rewriting is inferred by
  the rules: any rule of the form $\bnet{\alpha}{m}{-}{-}{X},\dots
  \Longrightarrow \bnet{\beta}{n}{-}{-}{X'}$ where $X \neq X'$ cannot
  safely be rewritten at the same time (or in the same \emph{pass}) as
  any rule that rewrites within $X$.
\item[Rewrite] For each active pair and agent within $\mathcal{C}_s$,
  apply rule.  For simple rules where rewriting does not occur within
  agent borders and there are no internal edges:
\begin{equation*}
\begin{split}
\bnetm{\alpha}{x}{u_1,\dots,u_m}, \bnetm{\beta}{x}{v_1,\dots,v_m} \Longrightarrow & \bnetm{\Gamma_1}{w_1}{w_2,\dots,w_p},\dots,\\
& \bnetm{\Gamma_q}{w_r}{w_{r+1},\dots,w_s},\\
& u_1-w_i,\dots,u_m-w_{i'},\\
& v_1-w_{i''},\dots,v_n-w_{i'''}
\end{split}
\end{equation*}
where $\Gamma^{\textbf{w}}$ are the (possibly empty) agents that are
produced on rewriting and $\textbf{w}$ are the intermediary labels
given to the wires of the produced net.  Note that the size of $
\textbf{w} $ is potentially larger than the number of ports to the
left hand side of the rule.  The cases for rules whose agents have
internal ports and rewriting occurs across borders incorporates a
richer programmatic syntax and the resulting operational calculus is
more complex.
\item[Tidy] If $w$ is a label within $\Gamma$ and there exists $u-w$ then 
substitute $w$ by $u$ within $\Gamma$ ($\Gamma[w/u]$).
\end{itemize}

\section{Conclusion}\label{sec:conc}
We have presented a new visual language generalising interaction nets
to incorporate features from bigraphs. Domains of application
include concurrent and reactive systems. Not only can binets model
these systems both graphically and textually, but they are also
directly implementable. We are currently working on the implementation
of an abstract machine for binets, inspired by the interaction net
machines defined in~\cite{PintoJS:seqcam}.


\bibliographystyle{eptcs}
\bibliography{TERMGRAPH13}

\begin{thebibliography}{10}
\providecommand{\bibitemdeclare}[2]{}
\providecommand{\surnamestart}{}
\providecommand{\surnameend}{}
\providecommand{\urlprefix}{Available at }
\providecommand{\url}[1]{\texttt{#1}}
\providecommand{\href}[2]{\texttt{#2}}
\providecommand{\urlalt}[2]{\href{#1}{#2}}
\providecommand{\doi}[1]{doi:\urlalt{http://dx.doi.org/#1}{#1}}
\providecommand{\bibinfo}[2]{#2}

\bibitemdeclare{inproceedings}{AlvesS:gracp}
\bibitem{AlvesS:gracp}
\bibinfo{author}{Sandra \surnamestart Alves\surnameend},
  \bibinfo{author}{Maribel \surnamestart Fern{\'a}ndez\surnameend} \&
  \bibinfo{author}{Ian \surnamestart Mackie\surnameend} (\bibinfo{year}{2011}):
  \emph{\bibinfo{title}{A new graphical calculus of proofs}}.
\newblock In \bibinfo{editor}{\surnamestart Echahed\surnameend}, editor: {\sl
  \bibinfo{booktitle}{Proceedings of TERMGRAPH 2011}},
  \bibinfo{publisher}{EPTCS}, pp. \bibinfo{pages}{69--84}.
\newblock \urlprefix\url{http://dx.doi.org/10.4204/EPTCS.48.8}.

\bibitemdeclare{inproceedings}{AndreiO:PORGY}
\bibitem{AndreiO:PORGY}
\bibinfo{author}{Oana \surnamestart Andrei\surnameend},
  \bibinfo{author}{Maribel \surnamestart Fern{\'a}ndez\surnameend},
  \bibinfo{author}{H{\'e}l{\`e}ne \surnamestart Kirchner\surnameend},
  \bibinfo{author}{Guy~Melan\c \surnamestart {c}on\surnameend},
  \bibinfo{author}{Olivier \surnamestart Namet\surnameend} \&
  \bibinfo{author}{Bruno \surnamestart Pinaud\surnameend}
  (\bibinfo{year}{2011}): \emph{\bibinfo{title}{PORGY: Strategy-Driven
  Interactive Transformation of Graphs}}.
\newblock In \bibinfo{editor}{\surnamestart Echahed\surnameend}, editor: {\sl
  \bibinfo{booktitle}{Proceedings of TERMGRAPH 2011}},
  \bibinfo{publisher}{EPTCS}, pp. \bibinfo{pages}{54--68}.
\newblock \urlprefix\url{http://dx.doi.org/10.4204/EPTCS.48.7}.

\bibitemdeclare{article}{AndreiK08}
\bibitem{AndreiK08}
\bibinfo{author}{Oana \surnamestart Andrei\surnameend} \&
  \bibinfo{author}{H{\'e}l{\`e}ne \surnamestart Kirchner\surnameend}
  (\bibinfo{year}{2008}): \emph{\bibinfo{title}{A Rewriting Calculus for
  Multigraphs with Ports}}.
\newblock {\sl \bibinfo{journal}{Electr. Notes Theor. Comput. Sci.}}
  \bibinfo{volume}{219}, pp. \bibinfo{pages}{67--82}.
\newblock \urlprefix\url{http://dx.doi.org/10.1016/j.entcs.2008.10.035}.

\bibitemdeclare{article}{AspertiA:bolhom}
\bibitem{AspertiA:bolhom}
\bibinfo{author}{Andrea \surnamestart Asperti\surnameend},
  \bibinfo{author}{Cecilia \surnamestart Giovannetti\surnameend} \&
  \bibinfo{author}{Andrea \surnamestart Naletto\surnameend}
  (\bibinfo{year}{1996}): \emph{\bibinfo{title}{The {B}ologna Optimal
  Higher-order Machine}}.
\newblock {\sl \bibinfo{journal}{Journal of Functional Programming}}
  \bibinfo{volume}{6}(\bibinfo{number}{6}), pp. \bibinfo{pages}{763--810}.
\newblock \urlprefix\url{http://dx.doi.org/10.1017/S0956796800001994}.

\bibitemdeclare{article}{rhoCalIGLP-I+II-2001}
\bibitem{rhoCalIGLP-I+II-2001}
\bibinfo{author}{Horatiu \surnamestart Cirstea\surnameend} \&
  \bibinfo{author}{Claude \surnamestart Kirchner\surnameend}
  (\bibinfo{year}{2001}): \emph{\bibinfo{title}{The rewriting calculus ---
  {Part~I {\em and} II}}}.
\newblock {\sl \bibinfo{journal}{Logic Journal of the Interest Group in Pure
  and Applied Logics}} \bibinfo{volume}{9}(\bibinfo{number}{3}), pp.
  \bibinfo{pages}{427--498}.
\newblock \urlprefix\url{http://dx.doi.org/10.1093/jigpal/9.3.339}.

\bibitemdeclare{inproceedings}{Drewes-Hoffmann-Plump:00}
\bibitem{Drewes-Hoffmann-Plump:00}
\bibinfo{author}{Frank \surnamestart Drewes\surnameend},
  \bibinfo{author}{Berthold \surnamestart Hoffmann\surnameend} \&
  \bibinfo{author}{Detlef \surnamestart Plump\surnameend}
  (\bibinfo{year}{2000}): \emph{\bibinfo{title}{Hierarchical Graph
  Transformation}}.
\newblock In \bibinfo{editor}{\surnamestart Tiuryn\surnameend}, editor: {\sl
  \bibinfo{booktitle}{Proc.\ Foundations of Software Science and Computation
  Structures (FOSSACS 2000)}}, {\sl \bibinfo{series}{Lecture Notes in Computer
  Science}} \bibinfo{volume}{1784}, pp. \bibinfo{pages}{98--113}.
\newblock \urlprefix\url{http://dx.doi.org/10.1007/3-540-46432-8_7}.

\bibitemdeclare{article}{FernandezM:inMCpa}
\bibitem{FernandezM:inMCpa}
\bibinfo{author}{M.~\surnamestart Fern\'{a}ndez\surnameend} \&
  \bibinfo{author}{L.~\surnamestart Khalil\surnameend} (\bibinfo{year}{2003}):
  \emph{\bibinfo{title}{Interaction nets with {M}c{C}arthy's amb: Properties
  and Applications}}.
\newblock {\sl \bibinfo{journal}{Nordic Journal of Computing}}
  \bibinfo{volume}{10}(\bibinfo{number}{2}), pp. \bibinfo{pages}{134--162}.
\newblock \urlprefix\url{http://dx.doi.org/10.1016/S1571-0661(05)80363-9}.

\bibitemdeclare{inproceedings}{FernandezM:terrgi}
\bibitem{FernandezM:terrgi}
\bibinfo{author}{M.~\surnamestart Fern\'andez\surnameend} \&
  \bibinfo{author}{I.~\surnamestart Mackie\surnameend} (\bibinfo{year}{1996}):
  \emph{\bibinfo{title}{From Term Rewriting to Generalised Interaction Nets}}.
\newblock In: {\sl \bibinfo{booktitle}{Proceedings of PLILP'96. Programming
  Languages: Implementations, Logics, and Programs}}, {\sl
  \bibinfo{series}{Lecture Notes in Computer Science}} \bibinfo{volume}{1140},
  \bibinfo{publisher}{Springer-Verlag}.
\newblock \urlprefix\url{http://dx.doi.org/10.1007/3-540-61756-6_94}.

\bibitemdeclare{article}{MackieIC:intntr}
\bibitem{MackieIC:intntr}
\bibinfo{author}{Maribel \surnamestart Fern{\'a}ndez\surnameend} \&
  \bibinfo{author}{Ian \surnamestart Mackie\surnameend} (\bibinfo{year}{1998}):
  \emph{\bibinfo{title}{Interaction Nets and Term Rewriting Systems}}.
\newblock {\sl \bibinfo{journal}{Theoretical Computer Science}}
  \bibinfo{volume}{190}(\bibinfo{number}{1}), pp. \bibinfo{pages}{3--39}.
\newblock \urlprefix\url{http://dx.doi.org/10.1016/S0304-3975(97)00082-0}.

\bibitemdeclare{article}{EXPRESS}
\bibitem{EXPRESS}
\bibinfo{author}{Maribel \surnamestart Fern{\'a}ndez\surnameend},
  \bibinfo{author}{Ian \surnamestart Mackie\surnameend} \&
  \bibinfo{author}{Fran\c \surnamestart {c}ois R{\'e}gis~Sinot\surnameend}
  (\bibinfo{year}{2006}): \emph{\bibinfo{title}{Interaction Nets vs. the {\it
  rho}-calculus: Introducing Bigraphical Nets}}.
\newblock {\sl \bibinfo{journal}{Electr. Notes Theor. Comput. Sci.}}
  \bibinfo{volume}{154}(\bibinfo{number}{3}), pp. \bibinfo{pages}{19--32}.
\newblock \urlprefix\url{http://dx.doi.org/10.1016/j.entcs.2006.05.004}.

\bibitemdeclare{inproceedings}{FleutotF:encoci}
\bibitem{FleutotF:encoci}
\bibinfo{author}{Fabien \surnamestart Fleutot\surnameend}
  (\bibinfo{year}{2004}): \emph{\bibinfo{title}{Encoding an Object Calculus
  into Interaction Nets}}.
\newblock In \bibinfo{editor}{M.~\surnamestart Fernandez\surnameend}, editor:
  {\sl \bibinfo{booktitle}{Proc. of the 2nd Int. Workshop on Term Graph
  Rewriting (TERMGRAPH 2004)}}, \bibinfo{series}{ENTCS},
  \bibinfo{address}{Rome}.
\newblock \urlprefix\url{http://dx.doi.org/10.1016/j.entcs.2005.03.024}.

\bibitemdeclare{inproceedings}{GonthierG:geoolr}
\bibitem{GonthierG:geoolr}
\bibinfo{author}{Georges \surnamestart Gonthier\surnameend},
  \bibinfo{author}{Mart{\'\i }n \surnamestart Abadi\surnameend} \&
  \bibinfo{author}{Jean-Jacques \surnamestart L{\'e}vy\surnameend}
  (\bibinfo{year}{1992}): \emph{\bibinfo{title}{The Geometry of Optimal Lambda
  Reduction}}.
\newblock In: {\sl \bibinfo{booktitle}{Proceedings of the 19th {ACM} Symposium
  on Principles of Programming Languages (POPL'92)}}, \bibinfo{publisher}{ACM
  Press}, pp. \bibinfo{pages}{15--26}.
\newblock \urlprefix\url{http://dx.doi.org/10.1145/143165.143172}.

\bibitemdeclare{inproceedings}{MackieIC:diagrams}
\bibitem{MackieIC:diagrams}
\bibinfo{author}{Abubakar \surnamestart Hassan\surnameend},
  \bibinfo{author}{Ian \surnamestart Mackie\surnameend} \&
  \bibinfo{author}{Jorge~Sousa \surnamestart Pinto\surnameend}
  (\bibinfo{year}{2008}): \emph{\bibinfo{title}{Visual Programming with
  Interaction Nets}}.
\newblock In \bibinfo{editor}{Gem \surnamestart Stapleton\surnameend},
  \bibinfo{editor}{John \surnamestart Howse\surnameend} \&
  \bibinfo{editor}{John \surnamestart Lee\surnameend}, editors: {\sl
  \bibinfo{booktitle}{Diagrams}}, {\sl \bibinfo{series}{Lecture Notes in
  Computer Science}} \bibinfo{volume}{5223}, \bibinfo{publisher}{Springer}, pp.
  \bibinfo{pages}{165--171}.
\newblock \urlprefix\url{http://dx.doi.org/10.1007/978-3-540-87730-1_17}.

\bibitemdeclare{misc}{jensen03bigraphs}
\bibitem{jensen03bigraphs}
\bibinfo{author}{O.~\surnamestart Jensen\surnameend} \&
  \bibinfo{author}{R.~\surnamestart Milner\surnameend} (\bibinfo{year}{2004}):
  \emph{\bibinfo{title}{Bigraphs and mobile processes (revised)}}.
\newblock \bibinfo{howpublished}{Technical Report 580, Computer Laboratory,
  University of Cambridge}.

\bibitemdeclare{inproceedings}{LafontY:intn}
\bibitem{LafontY:intn}
\bibinfo{author}{Yves \surnamestart Lafont\surnameend} (\bibinfo{year}{1990}):
  \emph{\bibinfo{title}{Interaction Nets}}.
\newblock In: {\sl \bibinfo{booktitle}{Proceedings of the 17th {ACM} Symposium
  on Principles of Programming Languages ({POPL}'90)}}, \bibinfo{publisher}{ACM
  Press}, pp. \bibinfo{pages}{95--108}.
\newblock \urlprefix\url{http://dx.doi.org/10.1145/96709.96718}.

\bibitemdeclare{incollection}{LafontY:fropni}
\bibitem{LafontY:fropni}
\bibinfo{author}{Yves \surnamestart Lafont\surnameend} (\bibinfo{year}{1995}):
  \emph{\bibinfo{title}{From Proof Nets to Interaction Nets}}.
\newblock In \bibinfo{editor}{J.-Y. \surnamestart Girard\surnameend},
  \bibinfo{editor}{Y.~\surnamestart Lafont\surnameend} \&
  \bibinfo{editor}{L.~\surnamestart Regnier\surnameend}, editors: {\sl
  \bibinfo{booktitle}{Advances in Linear Logic}}, {\sl \bibinfo{series}{London
  Mathematical Society Lecture Note Series}} \bibinfo{volume}{222},
  \bibinfo{publisher}{Cambridge University Press}, pp.
  \bibinfo{pages}{225--247}.
\newblock \urlprefix\url{http://dx.doi.org/10.1017/CBO9780511629150.012}.

\bibitemdeclare{phdthesis}{MackieIC:phd}
\bibitem{MackieIC:phd}
\bibinfo{author}{Ian \surnamestart Mackie\surnameend} (\bibinfo{year}{1994}):
  \emph{\bibinfo{title}{The Geometry of Implementation}}.
\newblock Ph.D. thesis, \bibinfo{school}{Department of Computing, Imperial
  College of Science, Technology and Medicine}.

\bibitemdeclare{inproceedings}{MackieIC:efflei}
\bibitem{MackieIC:efflei}
\bibinfo{author}{Ian \surnamestart Mackie\surnameend} (\bibinfo{year}{2004}):
  \emph{\bibinfo{title}{Efficient $\lambda $-evaluation with interaction
  nets}}.
\newblock In \bibinfo{editor}{V.~\surnamestart van Oostrom\surnameend}, editor:
  {\sl \bibinfo{booktitle}{Proc.\ 15th Int.\ Conference on Rewriting Techniques
  and Applications (RTA'04)}}, {\sl \bibinfo{series}{Lecture Notes in Computer
  Science}} \bibinfo{volume}{3091}, \bibinfo{publisher}{Springer-Verlag}, pp.
  \bibinfo{pages}{155--169}.
\newblock \urlprefix\url{http://dx.doi.org/10.1007/978-3-540-25979-4_11}.

\bibitemdeclare{inproceedings}{MackieIC:vcc}
\bibitem{MackieIC:vcc}
\bibinfo{author}{Ian \surnamestart Mackie\surnameend} (\bibinfo{year}{2009}):
  \emph{\bibinfo{title}{A rewriting paradigm for program and algorithm
  animation}}.
\newblock In: {\sl \bibinfo{booktitle}{VL/HCC}}, \bibinfo{publisher}{IEEE}, pp.
  \bibinfo{pages}{170--173}.
\newblock
  \urlprefix\url{http://doi.ieeecomputersociety.org/10.1109/VLHCC.2009.5295272%
}.

\bibitemdeclare{inproceedings}{MackieIC:tamc}
\bibitem{MackieIC:tamc}
\bibinfo{author}{Ian \surnamestart Mackie\surnameend} (\bibinfo{year}{2010}):
  \emph{\bibinfo{title}{A Visual Model of Computation}}.
\newblock In \bibinfo{editor}{Jan \surnamestart Kratochv\'{\i }l\surnameend},
  \bibinfo{editor}{Angsheng \surnamestart Li\surnameend},
  \bibinfo{editor}{Jir\'{\i } \surnamestart Fiala\surnameend} \&
  \bibinfo{editor}{Petr \surnamestart Kolman\surnameend}, editors: {\sl
  \bibinfo{booktitle}{TAMC}}, {\sl \bibinfo{series}{Lecture Notes in Computer
  Science}} \bibinfo{volume}{6108}, \bibinfo{publisher}{Springer}, pp.
  \bibinfo{pages}{350--360}.
\newblock \urlprefix\url{http://dx.doi.org/10.1007/978-3-642-13562-0_32}.

\bibitemdeclare{inproceedings}{Milner}
\bibitem{Milner}
\bibinfo{author}{Robin \surnamestart Milner\surnameend} (\bibinfo{year}{2001}):
  \emph{\bibinfo{title}{Bigraphical Reactive Systems}}.
\newblock In \bibinfo{editor}{Kim~Guldstrand \surnamestart Larsen\surnameend}
  \& \bibinfo{editor}{Mogens \surnamestart Nielsen\surnameend}, editors: {\sl
  \bibinfo{booktitle}{CONCUR}}, {\sl \bibinfo{series}{Lecture Notes in Computer
  Science}} \bibinfo{volume}{2154}, \bibinfo{publisher}{Springer}, pp.
  \bibinfo{pages}{16--35}.
\newblock \urlprefix\url{http://dx.doi.org/10.1007/3-540-44685-0_2}.

\bibitemdeclare{inproceedings}{PintoJS:seqcam}
\bibitem{PintoJS:seqcam}
\bibinfo{author}{Jorge~Sousa \surnamestart Pinto\surnameend}
  (\bibinfo{year}{2000}): \emph{\bibinfo{title}{Sequential and Concurrent
  Abstract Machines for Interaction Nets}}.
\newblock In \bibinfo{editor}{J.~\surnamestart Tiuryn\surnameend}, editor: {\sl
  \bibinfo{booktitle}{Proceedings of Foundations of Software Science and
  Computation Structures (FOSSACS)}}, {\sl \bibinfo{series}{Lecture Notes in
  Computer Science}} \bibinfo{volume}{1784},
  \bibinfo{publisher}{Springer-Verlag}, pp. \bibinfo{pages}{267--282}.
\newblock \urlprefix\url{http://dx.doi.org/10.1007/3-540-46432-8_18}.

\end{thebibliography}

\end{document}